\documentclass[prb,twocolumn,longbibliography]{revtex4-1}

\usepackage{amsmath}  
\usepackage{amsfonts} 
\usepackage{graphicx} 
\usepackage{subfigure}
\usepackage{epstopdf}
\usepackage{natbib}
\usepackage{float}
\usepackage[colorlinks=true,linkcolor=blue,citecolor=blue,urlcolor=blue]{hyperref}
\usepackage[capitalize]{cleveref}

\begin{document}


\title{Optical Non-Reciprocity in Coupled Resonators by Detailed Balance}
	\author{Yi-Xuan Yao}
	\author{Qing Ai}
	\email{aiqing@bnu.edu.cn}
	\affiliation{%
		Department of Physics, Applied Optics Beijing Area Major Laboratory, Beijing Normal University, Beijing 100875, China
	}%




\date{\today}

\begin{abstract}
	Inspired by the photosynthetic energy transfer process, we theoretically propose a method to realize non-reciprocal optical transmission in an array of coupled resonators. The optical non-reciprocity of the coupled resonators arises from the frequency gradient between adjacent cavities and the interaction with the environment, which is similar to photosynthetic energy transfer. An increase in the frequency gradient or the number of the cavities can lead to better non-reciprocity. However, although a higher environment temperature will increase the total photon number in the coupled cavities, non-reciprocity will be weakened. All these findings can be well described by the detailed balance. Our discovery reveals the similarity between the noise-induced optical non-reciprocity and exciton energy transfer in natural photosynthesis.
\end{abstract}

\maketitle 

\section{Introduction} 

Structures in living beings may have significant effects on physiological functions and thus enlighten inventions of bionic devices, e.g. honeycomb and light-harvesting antenna in photosynthesis. In natural photosynthesis, it generally includes three steps \cite{Fleming1994PT}. First of all, the solar photon is absorbed by the outer antenna. It is followed that the absorbed energy is delivered to the reaction center. Finally, the chemical reaction in which the energy is stored in the chemical products is activated. In the first two steps, the outer antenna is made up of pigments, e.g. chlorophylls and carotenes, and protein skeleton. The chlorophyll molecule consists of a chromphore, which can be modeled as a two-level atom, and a long tail \cite{Blankenship2014}. When the chlorophyll molecule absorbs the solar photon, the chromophore is excited from the ground state to the excited state. During the energy transfer process, the chromophore at the excited state will excite a nearby chromophore at the ground state and return to the ground state. In this way, the excitation state and thus the excitation energy is delivered from the outer antenna to the reaction center. In Fig.~\ref{fig:scheme}(a), the spatial arrangement of 7 choromophores in Fenna-Matthews-Olson (FMO) complex and energy transfer pathways are schematically illustrated \cite{Moix2011JPCL,Wu2012JCP,Ai2013JPCL}.
In order to make the energy transfer unidirectional towards the reaction center, energy gradient in the site energies of the chomophores is employed. 
Similar devices, e.g. funnel, have been widely used in daily life when pouring oil into a bottle in order to avoid waste, cf. Fig.~\ref{fig:scheme}(b).
 However, it is known that the energy difference between two states alone will not induce the irreversible and unidirectional transfer of population from one state to the other, as the quantum dynamics is coherent \cite{Scully1997}. Many studies have shown that the noise, i.e., the interaction with the environment, plays a prominent role in this process of energy transfer \cite{Mohseni2008JCP,Caruso2009JCP,Schmidt2011JPCL,Plenio2008NJP,Chin2010NJP}. 
It is the unavoidable environmental noise in combination with the energy gradient in the site energies that lead to an optimal energy transfer efficiency.




The coupled chromophores in photosynthesis form a network of chromophores and it is analogue to an array of coupled resonators.
A resonator localizes the eigen-modes of electromagnetic field through a ``mirror-cavity-mirror" structure \cite{Bliokh2008RMP}. In a closed resonator, the wavefunction of each mode $\Psi(\boldsymbol{r},t)=\psi(t)\xi (\boldsymbol{r})$ includes the time-domain $\psi(t)$, with the resonant frequency $\omega$, and the spatial structure of the field $\xi (\boldsymbol{r})$. Resonators can be coupled by the fields penetrating through the barriers and thus form the coupled resonators \cite{Bliokh2008RMP}, cf. Fig.~\ref{fig:scheme}(c). Interestingly, by adding one atom in one of the coupled resonators, perfect refraction can be switched on/off by tuning the detuning between the atomic transition frequency and the resonant frequency of the resonator \cite{Zhou2008PRL}.

The coupled resonators have been shown to demonstrate numerous interesting phenomena, e.g. quantum router \cite{Zhou2013PRL} and the non-reciprocal optical transmission \cite{Huang2021LSA,Zhu2021AdP}.
According to the Lorentz reciprocity theorem, the transmittance of the light will not be modified when the emission port and the reception port are exchanged \cite{Lorentz1896}. However, by magnetic-field-induced breaking of time-reversal symmetry, spatio-temporal modulation of system permittivity, and nonlinearity, optical non-reciprocity can be realized \cite{Jalas2013NP,Caloz2018PRA}. Non-reciprocal devices, e.g. optical isolators, optical circulators, and directional amplifiers, play a crucial role in communication
and quantum information processing. Interestingly, a levitated rotating nano-dimaond with NV centers can demonstrate non-reciprocal optical transmission because of electromagnetic-induced transparency  and Doppler effect \cite{Huang2022AdP}. 

In this paper, inspired by the rapid progress in photosynthetic energy transfer and optical non-reciprocity, we theoretically propose a method to realize non-reciprocal optical transmission in an array of coupled resonators. Analogue to the photosynthetic light harvesting, the optical non-reciprocity in the coupled resonators is achieved by gradient in the resonant frequencies of the resonators, cf. Fig.~\ref{fig:scheme}(d), and the system-bath interaction. Impressively, the side-coupled resonators with the enclosed synthetic magnatic flux can induce the non-reciprocal transport in the coupled-resonator array through introducing the environment induced gain or loss,\cite{Jin2016SciRep,Du2020SciRep,Jin2018PRL,Jin2018PRA} which is similar to the physical mechanism of non-reciprocity in this work. In contrast to the above three criteria \cite{Jalas2013NP,Caloz2018PRA}, the optical non-reciprocity in this work is due to breaking of time-reversal symmetry induced by the system-bath entanglement. 

This paper is organized as follows. In the next section, the model for optical non-reciprocal transmission and the quantum master equation approach is introduced. In Sec.~\ref{sec:Results}, the effects of the frequency gradient, the temperature and the number of coupled resonators on the non-reciprocal transmission are analyzed by the numerical simulations. Finally, the main discoveries have been summarized in Sec.~\ref{sec:Conclusion}.


\section{Model}\label{Sec:Model}

\begin{figure}[htbp]
	\centering
	\includegraphics[width=8cm]{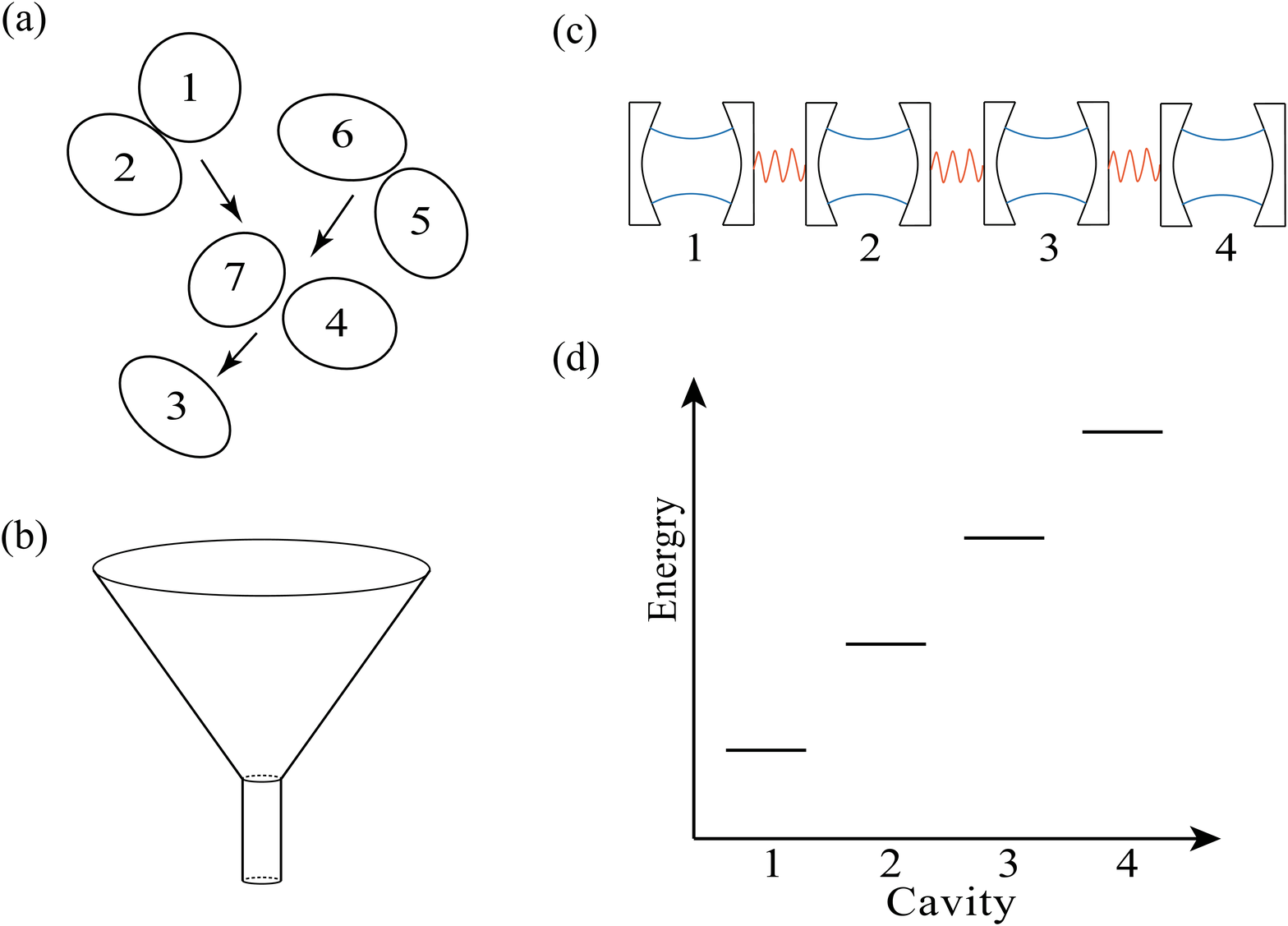}
	\caption{Schematic illustration of the similarity between excitation energy transfer in photosynthesis and artificial non-reciprocal transmission. (a) Spatial arrangement of chlorophylls and excitation energy transfer pathways in the monomeric subunit of the FMO complex, (b) illustration of an energy funnel, (c) structure of coupled cavities, (d) energy levels of the cavities.\label{fig:scheme}}
\end{figure}

As shown in Fig.~\ref{fig:scheme}(c), the Hamiltonian of the coupled resonators reads
\begin{equation}	H_S=\sum_{j}\omega_{j}a_{j}^{\dagger}a_{j}-\xi\sum_{j}\left(a_{j}^{\dagger}a_{j+1}+a_{j+1}^{\dagger}a_{j}\right),
\end{equation}
where $a_{j}^{\dagger}$ ($a_{j}$) is the creation (annihilation)
operator of the $j$th cavity with frequency $\omega_{j}$, $\xi$
is the coupling constant between two adjacent resonators, and we have
assumed $\hbar=1$. Here, we assume an energy gradient in the frequencies of resonators, as illustrated in Fig.~\ref{fig:scheme}(d).

When there are interactions with the environments, the open quantum
dynamics is determined by the quantum master equation \cite{Carmichael2009}
\begin{eqnarray}\label{eq:meq}
	\frac{d\rho}{dt} & = & -i[H_{S},\rho]+\mathcal{L}[\rho], \label{eq:me}\\
	\mathcal{L}[\rho] & = & \frac{\gamma}{2}\sum_{j}\big[(N_j^\textrm{th}+1)(2a_{j}\rho a_{j}^{\dagger}-a_{j}^{\dagger}a_{j}\rho-\rho a_{j}^{\dagger}a_{j})\nonumber \\
	&  & \quad\qquad+N_j^\textrm{th}(2a_{j}^{\dagger}\rho a_{j}-a_{j}a_{j}^{\dagger}\rho-\rho a_{j}a_{j}^{\dagger})\big],
	\label{Lrho}
\end{eqnarray}
where $\rho$ is the reduced density matrix of the coupled resonators, $N_j^\textrm{th}=(e ^{\omega_{j}/k_{B}T}-1)^{-1}$ is the average
photon number in the bath with $k_B$ and $T$ being the Boltzmann constant and the temperature respectively, $\gamma$ is the relaxation rate at zero temperature.
We use QuTiP \cite{Johansson2012CPC,Johansson2013CPC} to solve the quantum master equation numerically. In deriving the quantum master equation, several approximations have been applied, e.g. Born approximation, Markovian approximation, and secular approximation \cite{Breuer2007,Tao2020SB}. Generally speaking, the generators in the Lindblad operator (\ref{Lrho}) are the raising and lowering operators between the eigen states of the system Hamiltonian $H_S$ \cite{Breuer2007}. Here, we approximate them as the creation and annihilation operators of the bare cavities in order to better illustrate the underlying physical mechanism without being trapped by the sophisticated mathematics. We also remark that the exact open quantum dynamics can be reproduced by the hierarchical equation of motion and a recently-proposed quantum simulation approach in the high-temperature limit \cite{Buluta2009Science,Georgescu2014RMP,Wang2018NPJQI,Zhang2021FoP}.

\section{results}\label{sec:Results}

\begin{figure}[htbp]
	\centering
	\includegraphics[width=8.5cm]
	{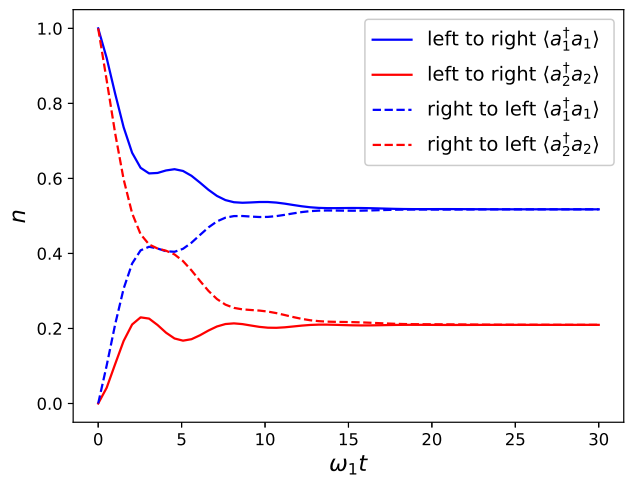}
	\caption{The non-reciprocal transmission in 2-coupled resonators with single photon incident from the left to the right (solid lines) and from the right to the left (dashed lines). The photon number in the first (second) cavity is labeled by the blue (red) lines. Here, $n_j=\langle a_j^\dagger a_j\rangle$ is the average photon number in the $j$th cavity. The parameters used in the calculation are $\omega_2=2\omega_1$, $\gamma=\xi=0.3\omega_1$, and $k_BT=\omega_1$.}\label{fig:2ce12}
\end{figure}

First of all, we adopt the system containing two cavities as a demonstration, and the eigenfrequency of the first cavity is lower than the second. Figure~\ref{fig:2ce12} shows the results of the case with $\omega_2=2\omega_1$. If we place one photon in the first cavity at the initial moment, the photon number in the first cavity $n_1$ will decline as depicted by the blue solid line and the photon number in the second cavity $n_2$ will rise as labeled by the red solid line, with $n_j=\langle a_j^\dagger a_j\rangle$ being the average photon number in the $j$th cavity. In this case, these two lines will not intersect before they reach their respective steady states. Moreover, at the short-time regime, they oscillate with a small amplitude due to the coherent term in the master equation~(\ref{eq:me}) and the small inter-resonator coupling in $H_S$ with respect to the large energy gap between the frequencies of two cavities. If we place one photon in the second cavity at the initial moment, the photon number in the first cavity $n_1$ will rise as denoted by the blue dashed line, and the photon number in the second cavity $n_2$ will decline as shown by the red dashed line. In this case, these two lines will intersect and reach the same steady states as in the previous case. Here, the photon number is always distributed more in the cavity with lower frequency no matter what the initial state is. The ratio between the photon numbers in the first and second cavities reads $\exp{(-\Delta\omega/k_BT)}$ with $\Delta\omega=\omega_2-\omega_1$ being the detuning between the two cavities, as determined by the detailed balance \cite{Mukamel1995,Breuer2007}.

\begin{figure}[htbp]
	\centering
	\includegraphics[width=8cm]
	{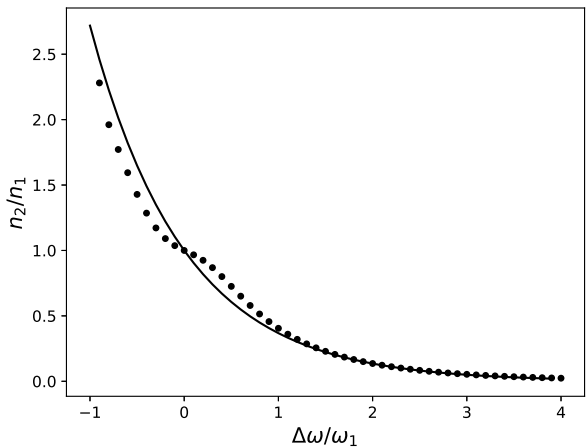}
	\caption{The ratio between the average photon numbers in the second and the first cavities $n_2/n_1$ as a function of frequency gradient $\Delta \omega$. The dots are obtained by the quantum master equation (\ref{eq:me}), while the solid line is obtained by Eq.~(\ref{eq:RhoRatio}). Other parameters are the same as~Fig.~\ref{fig:2ce12}.}
	\label{fig:nihe}
\end{figure}

In open quantum systems, the detailed balance subtly governs the instantaneous open quantum dynamics as well as the thermal equilibrium at the steady state. In order to explicitly reveal its effect, we resort to the quantum master equation (\ref{eq:me}). Notice that the Hamiltonian $H_S$ conserves the total number of the photons in all cavities. For the sake of simplicity, we restrict our consideration in the subspace with no more than one photon in either cavity, i.e., $a_j^\dagger\vert0\rangle$ ($j=1,2$) and $\vert0\rangle$ with $\vert0\rangle$ being the vacuum state for all cavities, and thus obtain
\begin{eqnarray}
\dot{\rho}_{00}&=&\gamma\sum_j[(N^\textrm{th}_j+1)\rho_{jj}- N^\textrm{th}_j\rho_{00}],\label{eq:rho00}\\
\dot{\rho}_{jj}&=&-\gamma(N^\textrm{th}_j+1)\rho_{jj}+\gamma N^\textrm{th}_j\rho_{00},\label{eq:rhojj}
\end{eqnarray}
where $\rho_{jj}=\langle0\vert a_j\rho a_j^\dagger\vert0\rangle$ is the probability of one photon in the $j$th cavity, $\rho_{00}=\langle0\vert \rho \vert0\rangle$ is the probability of all cavities in the vacuum state. In Eq.~(\ref{eq:rho00},\ref{eq:rhojj}) the coherent terms are neglected because of the large detuning condition $\Delta \omega \gg \xi$. 
At the steady state, since $\dot{\rho}_{jj}=\dot{\rho}_{00}=0$, we have $\rho_{jj}/\rho_{00}=N^\textrm{th}_j/(N^\textrm{th}_j+1)=\exp(-\omega_j/k_BT)$.
In the case of two cavities, the ratio between the probabilities of one photon in the two cavities reads
\begin{eqnarray}
\rho_{22}/\rho_{11}=\exp(-\Delta\omega/k_BT).\label{eq:RhoRatio}
\end{eqnarray}
In Fig.~\ref{fig:nihe}, we compare $n_{2}/n_{1}$ calculated by Eq.~(\ref{eq:me}) with $\rho_{22}/\rho_{11}$ by Eq.~(\ref{eq:RhoRatio}). It can be seen that the approximated analytic result reproduces the exact solution to the quantum master equation fairly well in the whole parameter regime, especially $\Delta\omega=0$ and $\Delta\omega\gg k_BT$. The small difference between them in the intermediate regime may be attributed to the truncation of three states in the above over-simplified model.

Transmission efficiencies denoted by the photon number are different in the cases of left-to-right and right-to-left propagation. Note that in the left-to-right propagation case, the photon number in the cavity at the exit end is $n_2$, i.e., the red solid line, while in the right-to-left propagation case, the photon number in the cavity at the exit end is $n_1$, i.e., the blue dashed line, which is higher than the former. In other words, the transmission of photons in the coupled-cavities array is non-reciprocal.

\begin{figure}[htbp]
	\centering
	\includegraphics[width=8.5cm]
	{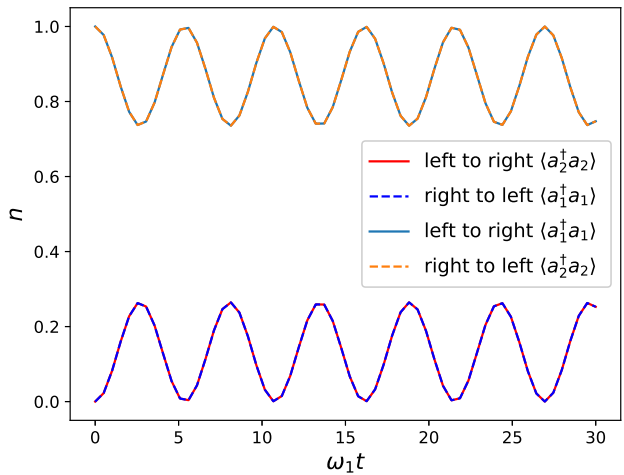}
	\caption{The transmission in 2-coupled resonators when $\gamma=0$.
		Other parameters are the same as Fig.~\ref{fig:2ce12}.
		When there is no interaction with the environment,
		the non-reciprocal transmission disappears.}
	\label{fig:2ce12_0-30gamma0}
\end{figure}

We further investigate the effect of the noise on the non-reciprocity. In Eq.~(\ref{Lrho}), when $\gamma = 0$, i.e., without the interaction with the environment, the time evolution of the density matrix is equivalent to that of an isolated system. It can be seen in Fig.~\ref{fig:2ce12_0-30gamma0} that without dissipation the photon number in the cavity at the incident end coherently oscillates around the initial value 1, while the photon number in the cavity at the exit end oscillates around 0. The sum of the photon numbers in both cavities is always unity, which is required by the conservation of probability. The small amplitude of oscillations is due to the relatively-large detuning, as denoted by $\xi<\Delta\omega$ in $H_S$. The red solid line and the blue dashed line are completely coincident with each other, which implies that the system is reciprocal.

To quantitatively evaluate the non-reciprocity of the system,
we define the transmission contrast as \cite{Caloz2018PRA}
\begin{equation}
	\eta=\frac{\lvert T_{+}-T_{-} \rvert}{T_{+}+T_{-}}=\frac{\lvert n_{+}-n_{-} \rvert}{n_{+}+n_{-}},
\end{equation}
where $T_{+}$ ($T_{-}$) is forward (backward) transmission coefficient, $n_{+}$ ($n_{-}$) is the steady-state photon number in the cavity at the exit end in the case of forward (backward) transmission. Here, $\eta=1$ implies the maximum non-reciprocity because the optical transmission is allowed unidirectionally, while $\eta=0$ suggests no non-reciprocity, as there is no difference between the optical transmissions in both directions.

\begin{figure}[htbp]
	\centering
	\includegraphics[width=8.5cm]
	{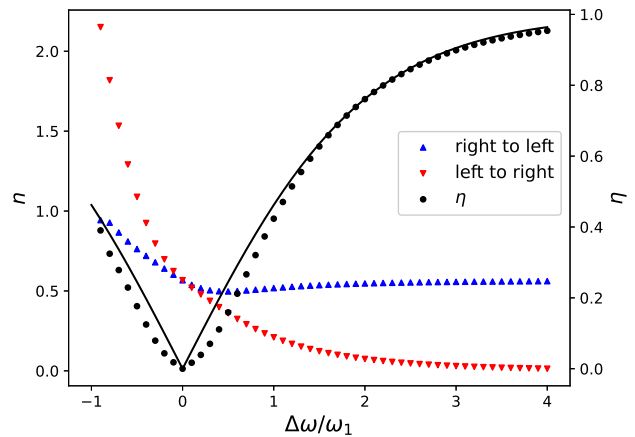}
	\caption{The photon number $n$ in the terminal cavity at steady state and the transmission contrast $\eta$ as a function of frequency gradient $\Delta \omega /\omega_1$ in the case of 2-coupled resonators. Other parameters are the same as~Fig.~\ref{fig:2ce12}.}
	\label{fig:w2xi0_3ga0_3}
\end{figure}

\begin{figure}[htbp]
	\centering
	\includegraphics[width=8.5cm]
	{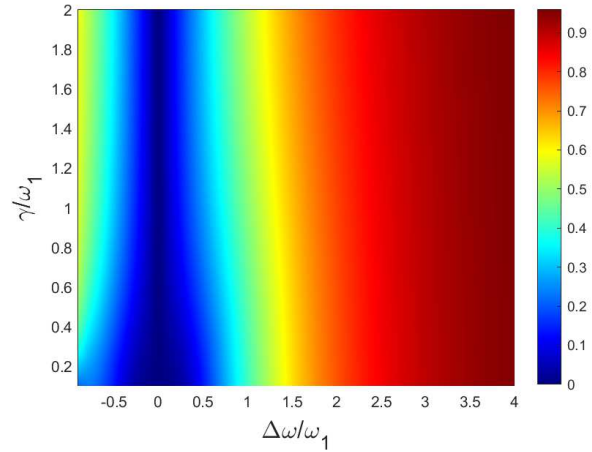}
	\caption{The dependence of the transmission contrast $\eta$ on the frequency gradient $\Delta \omega /\omega_1$ and the relaxation rate $\gamma$. Other parameters are the same as~Fig.~\ref{fig:2ce12}.}
	\label{fig:eta}
\end{figure}

We investigate the effect of the frequency gradient $\Delta\omega$ on the transmission contrast $\eta$ in Fig.~\ref{fig:w2xi0_3ga0_3}. When there is no difference between the frequencies of the two cavities, the transmissions in both directions are the same and thus reciprocal. If we gradually increase $\Delta\omega$, the change of the transmission from the right to the left is negligibly small, while the transmission from the left to the right decreases dramatically. In other words, the transmission becomes more and more non-reciprocal as $\Delta\omega$ is enlarged. Alternatively, we calculate the transmission contrast $\eta$ by the detailed balance  \cite{Mukamel1995,Breuer2007}, i.e.,
\begin{equation}
\eta=\frac{\lvert \frac{n_{+}}{n_{-}}-1 \rvert}{\frac{n_{+}}{n_{-}}+1}=\frac{\lvert e^{\Delta \omega/k_BT}-1 \rvert}{e^{\Delta \omega/k_BT}+1}.
\end{equation}
Similar to Fig.~\ref{fig:nihe}, the results by both methods coincide with each other in Fig.~\ref{fig:w2xi0_3ga0_3}. Therefore, we may safely arrive at the conclusion that the detailed balance accounts for the non-reciprocal transmission observed in the coupled resonators. The black solid line in Fig.~\ref{fig:w2xi0_3ga0_3} is symmetric with respect to 0, while the black dotted line is not strictly symmetric with respect to 0. The asymmetry of the black dotted line reflects the effect of the states with more than one photon in the resonators and the coherence between the states. Furthermore, we investigate the dependence of $\eta$ on both $\Delta\omega$ and $\gamma$ in Fig.~\ref{fig:eta}. The transmission contrast $\eta$ is generally not influenced by $\gamma$, which is consistent with the prediction by the detailed balance. The non-reciprocal transmission can be effectively tuned by the frequency gradient.


\begin{figure}[htbp]
	\centering
	\includegraphics[width=8.5cm]
	{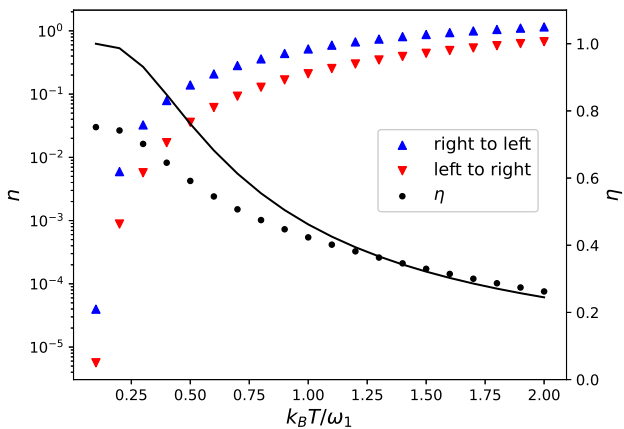}
	\caption{The photon number $n$ in the terminal cavity at steady state and the transmission contrast $\eta$ as a function of the temperature $T$ in the case of 2-coupled resonators. Other parameters are the same as~Fig.~\ref{fig:2ce12}.}
	\label{fig:2cT}
\end{figure}

In the above discussions, we assume that the environment is in the thermal equilibrium with a fixed temperature $T=\omega_1/k_B$. The effect of temperature on transmission is next observed in Fig.~\ref{fig:2cT}. It can be seen that when the temperature is absolute zero, there are no photons in the cavities at the final state because the photons initially in the cavity leaks to the environment. Despite the small $n$, the transmission contrast is close to unity and the non-reciprocity occurs. We remark that this non-reciprocal transmission is unfavorable, as the transmission has been greatly depressed as compared to the input. As the temperature increases, in both cases, the photon number in the terminal cavity increases. When the temperature is high to a certain extent, the sum of the photon number in the coupled cavities will be greater than 1 due to the heating effect of the environment. However, as the temperature increases, the transmission contrast $\eta$ decreases and the non-reciprocity eventually disappears for $k_BT\gg\Delta\omega$. To conclude, there exists an optimal temperature for the non-reciprocity at which both $n$ and $\eta$ are of the order of unity, i.e., $k_BT\sim\omega_1$.

\begin{figure}[htbp]
	\centering
	\includegraphics[width=8.5cm]
	{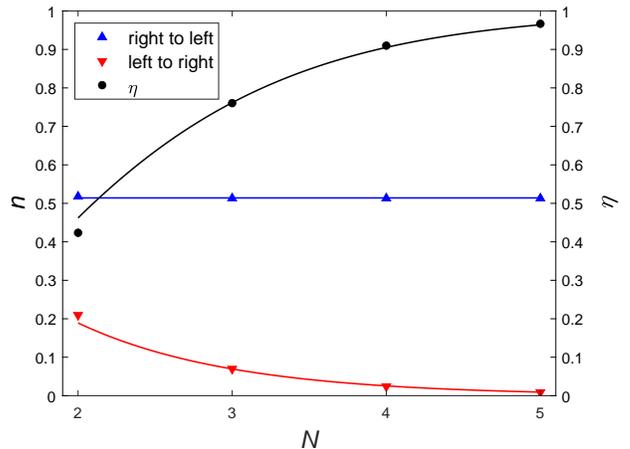}
	\caption{The photon number $n$ in the terminal cavity at steady state as a function of the number of cavities $N$, when the frequency gradient equals to $\omega_1$. The dots are obtained by solving the quantum master equation, while the solid lines are obtained by the detailed balance. Other parameters are the same as Fig.~\ref{fig:2ce12}.}
	\label{fig:cavnum}
\end{figure}

We further study the case with more cavities. We keep the frequency gradient of adjacent cavities $\omega_{j+1}-\omega_{j}$ and the eigenfrequency of the leftmost cavity $\omega_{1}$ constant and only vary the number of coupled cavities $N$. In~Fig.~\ref{fig:cavnum}, it can be seen that as the number of cavities $N$ increases, although it does not influence the photon number in the leftmost cavity, the photon number in the cavity at the right end decreases significantly. This is because more cavities will enlarge the gap between the eigenfrequencies of the cavities at both ends. As a result, the non-reciprocal transmission is enhanced for a larger $N$.

\section{Conclusion}\label{sec:Conclusion}

In this paper, we discuss the non-reciprocal optical transmission in an array of coupled resonators. We show that the optical non-reciprocity is akin to the photosynthetic energy transfer. In both cases, the non-reciprocal transmissions result from the frequency (energy) gradient and the interaction with environment. The open quantum dynamics and thus the non-reciprocity are analyzed by numerically solving the quantum master equation. It is demonstrated that an increase in the frequency gradient leads to better non-reciprocity. We also observe that there exists an optimal temperature at which both the transmission contrast and the absolute transmission are sufficiently high to ensure a favorable non-reciprocity. When the frequency gradient of adjacent cavities and the eigenfrequency of the first cavity are kept constant, the more cavities coupled, the better non-reciprocity of the system is.

In the quantum master equation~(\ref{eq:me}), we assume a time-independent relaxation rate $\gamma$. However, if the Markovian approximation is canceled, we would expect the non-Markovian effects might emerge \cite{Ishizaki2010PCCP,Caruso2010PRA}. For example, the entanglement and the quantum mutual information may not decay monotonically. Since the array of coupled resonators involves a network of multi-resonators, it may lose information to the environment in a cavity while it regain information from the environment in another cavity \cite{Chen2022NPJQI}.

To summarize, our findings bridge the gap between the non-reciprocal optical transmission and the energy transfer in photosynthesis. And it may enrich the understanding of the underlying physical mechanisms for both issues.

\begin{acknowledgments}
We are grateful to Ke-Yu Xia for drawing our attention to the similarity between the optical non-reciprocity and photosynthetic energy transfer. We thank stimulating discussions with Jun-Jie Lin and Wan-Ting He. This work is supported by Beijing Natural Science Foundation under Grant No.~1202017 and the National Natural Science Foundation of China under Grant Nos.~11674033, 11505007, and Beijing Normal University under Grant No.~2022129.
\end{acknowledgments}


%
	
\end{document}